# A shell model mass formula for exotic light nuclei

Mariano Bauer*, Hugo García Tecocoatzi** and Cristian Mojica**

*Instituto de Física, **Instituto de Ciencias Nucleares

Universidad Nacional Autónoma de México

México, D.F., 01000, MEXICO

**Abstract.** An analytic phenomenological shell model mass formula for light nuclei is constructed., The formula takes into account the non locality of the self consistent single particle potential and the special features of light nuclei, namely: a) charge and mass distributions are closer to a Gaussian shape than to the shape characteristic in medium and heavy nuclei; b) the central charge and mass densities are larger than, and decrease towards, the "asymptotic" values that are the reference parameters for nuclear matter; and c) after a shell closure, the next level has a larger orbital angular momentum and a noticeably larger mean square radius. Only then a good numerical fit is obtained with parameters consistent with optical model analysis and empirical spin-orbit couplings. A correlation between the "skin effect" and the symmetry dependence of the optical potential is established. Towards the neutron drip line the potential well depth, the spin-orbit splitting of the single particle levels and the gap between major shells decrease, as has been observed. The ensuing shift and contraction of the single particle level scheme may lead to: a) to strong configuration mixing and new magic numbers, and b) the onset of the halo effect, to avoid the expulsion of single particle levels to the continuum.

1. Introduction

Experiments have extended the scope of known nuclei to about 3000, reaching close to the drip lines in the region of light nuclei. As a consequence, new features have been discovered, like halo and skin nuclei, dissolution of magic numbers, new magic numbers [1-11]. Whether these new features are present in the mass formulae fitted to nuclei close to the stability line is an area of current research [12-14]. Bethe-Weizsacker semi empirical mass formulae based on the liquid drop model, together with the addition of surface diffuseness effects and of shell model and pairing corrections based on a Fermi model, have provided excellent fits to the extended nuclear mass surfaces [15]. It is however recognized that the larger differences with empirical data lie in the light nuclei region.

In the present work, the phenomenological shell model approach, successfully applied to medium and heavy nuclei [16, 17], is used to calculate binding energies of light nuclei. The results are shown to depend substantially upon taking into account the way the nuclear charge and mass distributions of light nuclei differ from those of the medium and heavy nuclei that give rise to the liquid drop model. Presence of the above mentioned new features is recognized in the expressions derived.

## 2. The distinctive features of light nuclei

The analytic phenomenological shell model for light nuclei must take into account the following features:

a) The charge and mass distributions for light nuclei are closer to a Gaussian shape than to the Fermi or Woods-Saxon shapes characteristic in medium heavy and heavy nuclei [22].
b) The central charge and mass densities for light nuclei are larger than, and decrease towards, the "asymptotic" values that set in already for medium heavy nuclei and become the reference parameters for nuclear matter. The ratio of the light nuclei central density to the asymptotic value is of the order of 1.2 -1.3 [22]. The self consistent potential well depths are expected to follow this trend.
c) After a shell closure, the next level that begins to be filled has a larger orbital angular momentum and consequently a noticeably larger mean square radius. The nuclear radius is expected to reflect this. Thus, in addition to the overall $A^{1/3}$ dependence usually considered, a discontinuous increase is expected to occur after crossing a magic number (Note that this is not expected to occur for heavier nuclei where the spin orbit splitting locates the sublevels $j = l + ½$ and $j = l - ½$ in consecutive major shells).

Following a), a Gaussian shape is assumed for the mass density distribution of light nuclei, namely:

$$\rho(r) = \rho_0 \exp(-r^2/b^2) \quad (1)$$

Integrating over volume, the total number of nucleons $A = N+Z$ is equal to $A = \rho_0 \pi^{3/2} b^3$ so that

$$b = (\rho_0 \pi^{3/2})^{1/3} A^{1/3} = b_o A^{1/3} \quad (2)$$

Thus the range of the Gaussian distribution has the same A-dependence as the range of the uniform distribution $R = r_0 A^{1/3}$ usually assumed for medium and heavy nuclei, requiring naturally that the product $\rho_0 b_o^3$ remains independent of A..

An interesting consequence is this. As the self consistent potential well depth $W_0$ is proportional to the central density, one expects the product $W_0 b_o^3$ also to remain independent of A. To maintain this, any additional A-dependence in one of these parameters to take into account feature b), that affects the well depth ($W_0 \to W_0(A, Z)$), and feature c), that affects the mass density range ($b_0 \to b_0(A, Z)$), needs a compensating effect in the other. For example, the symmetry dependence of the potential well depth derived from optical model fits in a unified description of bound and scattering states [20, 21], namely

$$V(N, Z) = V_0 + \tau V_1 (N-Z)/A = V_0 \{1+ \tau (V_1/V_0)(N-Z)/A\} \qquad (3)$$

where $\tau = 1$ for protons and $\tau = -1$ for neutrons, has to be taken into account in the mass distribution range $b_o$ by a compensating factor

$$\{1+ \tau (V1/V0) (N-Z)/A\}^{-1/3} \sim \{1- \tau (1/3) (V1/V0) (N-Z)/A + \ldots\}$$

Accordingly, the range increases both for neutron or proton excess, giving rise to a "skin" effect, as has been observed [5]. It is thus seen that the skin effect is linked to the asymmetry term required in optical model fits, or vice versa, both as manifestations of the isospin dependence of the nucleon-nucleon interaction.

Similar compensating effects have to be introduced when features b) and c) are parameterized.

### 3. The phenomenological shell model

The shell model of nuclei assumes that nucleons move independently in a Hartree-Fock self-consistent potential that essentially follows the mass distribution radial dependence, is necessarily non local due to the antisymetrization required by the Pauli principle and contains a spin-orbit interaction to yield the correct magic numbers. Residual interactions tend to pair nucleon total angular momenta, preserving sphericity near magic numbers and giving rise to the pairing energy that separates the nuclear mass surfaces [18].

For an even-even nucleus with N neutrons and Z protons, without residual interactions, the total energy is taken as a sum of independent particle systems contributions, in addition to a Coulomb energy, namely:

$$E(N, Z) = E(N) + E(Z) + EC(Z) \qquad (4)$$

with:

$$EC(Z) = a_C \{Z(Z-1)/2\} / (b_0(A)/b_0) A^{1/3} \qquad (5)$$

and

$$E(N) = \Sigma_{nljm} \{e_{nljm} - \tfrac{1}{2} <W(r)>_{nljm} \} = \Sigma_{nljm} E_{nljm} \qquad (6)$$

and similarly for E(Z). The $e_{nljm}$ are the single particle energies in the self consistent non local potential for a spherical nucleus, and $<W(r)>_{nljm}$ the corresponding potential energies. The binding energy is then B(N,Z) = - E(N,Z).

In accordance with feature a), a simple harmonic oscillator potential plus spin orbit coupling is adopted, that is, without the Nilsson term $-Dl^2$ used to correct for the effect of a flatter Woods-Saxon shape:

$$W(r) = - W_0 + \tfrac{1}{2} m\omega r^2 + C \, \mathbf{l.s} \qquad (7)$$

The effective mass approximation treatment of the non locality in the self consistent potential (details in [19]. and in Appendix A) yields:

$E(N) = E(\text{closed shells}) + E(\text{partially filled shell}) =$

$$= 1/3 \, (\eta+1)(\eta+2)(\eta+3)[ - \tfrac{1}{2} W_0 + \tfrac{3}{4} \mathcal{A} (\eta+2) - (9/10) \mathcal{B} \{(\eta+2)^2 + 1\}] \qquad (8)$$

$+ \alpha (\eta+2)(\eta+3)[ - \tfrac{1}{2} W_0 + \mathcal{A}(\eta+5/2) - (3/2) \mathcal{B} \{(\eta+2)^2 + (\eta+2) + 1\}] f(\alpha,\eta)$

+ the spin orbit contribution in the partially filled shell ( Eqs. A 9, A9*)

with

$$\mathcal{A} = \tfrac{1}{4} [2+(m^*/m)] \, \hbar\omega^* \qquad \mathcal{B} = (1/16) [1-(m^*/m)] (\hbar\omega^*)^2/W_0$$

$$\hbar\omega^* = \hbar\omega \, (m/m^*)^{1/2} \qquad f(\alpha,\eta) = - \{1.1 - 0.15 \, \alpha \, (1 + 0.7 \, \eta)\} \qquad (9)$$

Here $\eta$ denotes the principal quantum number of the last filled neutron shell. The number of neutrons in closed shells is $(1/3)(\eta+1)(\eta+2)(\eta+3)$ and the number of neutrons in the partially filled $\eta+1$ shell is $\alpha(\eta+2)(\eta+3)$ with $0 \leq \alpha \leq 1$. The spin-orbit coupling has been assumed to be proportional to $(1/r)(dW(r)/dr)$, yielding for a harmonic oscillator potential a coefficient

$$C = \lambda \, (\hbar\omega^*)^2/2mc^2 \qquad (9')$$

where $\lambda$ is an adjustable parameter.

The effective mass m* < m is related to the range of the non locality and is taken as the parameter representing this feature. The fact that m* is smaller than m results in a larger oscillator frequency and a larger potential depth (see Eqs.(9) and (10)). This leads to larger binding energies than those that would arise from a purely local potential and allows to attain the experimental values [17, 19].

One has an equivalent expression for E(Z) with the corresponding proton parameters.

Although summed separately, the neutron and proton contributions are not entirely independent. Following the unified description of bound and scattering single particle states [20,21], one has to take into account the relation in potential well depths expected from optical model analysis of nucleon scattering, namely:

$$W_0^{n,p} = (\frac{m}{m*})_{n,p} \quad V_{n,p} = (\frac{m}{m*})_{n,p} V_0 \{1 + \tau(V_1/V_0)(N-Z)/A\} \tag{10}$$

with $\tau = 1$ for protons and $\tau = -1$ for neutrons. This asymmetry dependence is reflected also in the corresponding effective masses as:

$$\left(\frac{m}{m*}\right)_{n,p} = \left(\frac{m}{m*}\right)_{av} \{1 + \tau((\frac{m}{m*})_{av} - 1)(V_1/V_0)(N-Z)/A\} \tag{11}$$

where $\left(\frac{m}{m*}\right)_{av} = \frac{1}{2}\{\left(\frac{m}{m*}\right)_n + \left(\frac{m}{m*}\right)_p\}$; as well as in the oscillator frequencies that are given by [20]:

$$(\hbar\omega^*)_{n,p} = (3/2)^{\frac{1}{3}} \frac{(\hbar c)^2}{2mc^2} \left(\frac{m}{m*}\right)_{n,p} (1/b_0)^2 A^{-\frac{1}{3}} \tag{12}$$

Finally, the special features of light nuclei are parameterized as follows. The decrease of the potential well depth of about 20% between Be and Ca is taken as

$$V_0(A) = V_0\{1 - [a_1 - a_2 \exp(-0.04(A-4)^2)] - \frac{a_3\tau(N-Z)}{A}\} \tag{13}$$

The increase in potential range on crossing a magic number is taken into account by multiplying the range by an extra factor (Appendix B):

$$\{1 + a_4\alpha_{n,p}\eta_{n,p}\} \tag{14}$$

To include the compensating effects needed to maintain $V_0(A)b_0(A)^3$ independent of A, expression (13) is multiplied by $\{1 + a_4\alpha_{n,p}\eta_{n,p}\}^{-3}$ and expression (14) by

$$\{1 - [a_1 - a_2 \exp(-0.04(A-4)^2)] - \frac{a_3\tau(N-Z)}{A}\}^{-1/3} \quad .$$

4. Results

*4.1 Binding energies*

To test the model, even-even nuclei with $4 \leq N, Z \leq 20$ are considered, covering the filling of the 1p and the (2s, 2d) shells, as characterized by the principal and orbital angular momentum quantum numbers. For the other nuclei the binding energies require the addition of pairing terms, that can be derived from the superfluid model [18] as:

- $-\Delta_n$      for odd N even Z
- $-\Delta_p$      for even N odd Z
- $-\Delta_n - \Delta_p$      for odd N odd Z,

where $\Delta_n$ and $\Delta_p$ are the corresponding energy gaps, giving rise to four nuclear surfaces.

As here constructed, the total energy expression has ten adjustable parameters: $V_0$, $V_1$, $b_0$, $\left(\frac{m}{m*}\right)_{av}$, $a_C$ and $\lambda$, in addition to those introduced to take into account the special features b) and c), namely $a_1$ to $a_4$. To be noted however is that they are not completely free as they must fall in the range of values arising from optical model elastic scattering fits, experimental sizes and spin-orbit splittings, etc. Consequently the purpose is to illustrate, if possible, that a reasonable fit can be obtained with parameter values congruent with other empirical evidence. Thus, no mean square root fit was attempted to achieve an optimal agreement, as this analysis leaves room for additional, although small, differences between neutron and proton parameters [25] or the possibility of configuration mixing due to residual interactions.

To begin with, the values $a_1 = 0.21$, $a_2 = 0.166$ and $a_3 = 0.869$ yield a 21 % decrease in potential well depth from $^4$He to $^{40}$Ca; and $a_4 = 0.145$ assures a 14.5% increase in range as the (2s, 2d) shell is filled, satisfying expected values of features b) and c). Following this, the set:

$V_0$= 67.5 MeV, $V_1$= 25 MeV, $b_0$= 0.805 F, (m/m*) = 1.05  $a_C$ = 0.84 MeV, $\lambda$ = 14

yields the results shown in Table 1. These parameters are in the range of values of extensive optical model fits, such as in ref 24.

The rms deviation of total binding energies with respect to the 56 experimental values known [26] is large, 11.40 MeV, arising mostly from nuclei away from the valley of stability. An intermediate set of only 32 nuclei close to the stability line yields already an rms deviation of 3.56 MeV. Finally, for the fourteen stable nuclei with N and Z even, the rms deviation is only 1.26 MeV. In this last case, the largest deviations arise from $^{24}$Mg and $^{28}$Si; removal of these reduces the rms to 0.60 MeV.

| N = | 4 | 6 | 8 | 10 | 12 | 14 | 16 | 18 | 20 |
|---|---|---|---|---|---|---|---|---|---|
| Z = | | | | | | | | | |
| 4 | **6.832** | **6.701** | **5.682** | **3.696** | **2.543** | **1.839** | **1.412** | **1.073** | **0.914** |
| | *7.062* | *6.497* | *5.720* | *4.994* | *4.270* | *--* | *--* | *--* | *--* |
| 6 | **6.178** | **7.760** | **7.739** | **6.747** | **5.917** | **5.302** | **4.482** | **4.412** | **4.126** |
| | *6.032* | *7.680* | *7.520* | *6.922* | *6.425* | *5.950* | *--* | *--* | *--* |
| 8 | **4.724** | **7.250** | **7.969** | **7.809** | **7.375** | **6.977** | **6.625** | **6.245** | **5.963** |
| | *4.879* | *7.052* | *7.976* | *7.767* | *7.568* | *7.365* | *7.016* | *6.457* | *5.925* |
| 10 | **2.402** | **5.855** | **7.379** | **8.065** | **8.142** | **8.085** | **7.961** | **7.738** | **7.754** |
| | *--* | *6.082* | *7.741* | *8.032* | *8.080* | *7.992* | *7.753* | *7.390* | *7.040* |
| 12 | **0.959** | **4.672** | **6.557** | **7.742** | **8.149** | **8.335** | **8.389** | **8.303** | **8.217** |
| | *--* | *--* | *6.723* | *7.662* | *8.260* | *8.332* | *8.272* | *8.055* | *7.807* |
| 14 | **-0.015** | **3.738** | **5.806** | **7.316** | **7.958** | **8.330** | **8.531** | **8.564** | **8.567** |
| | *--* | *--* | *--* | *7.166* | *7.924* | *8.447* | *8.520* | *8.481* | *8.336* |
| 16 | **-0.705** | **2.980** | **5.129** | **6.849** | **7.661** | **8.175** | **8.493** | **8.627** | **8.707** |
| | *--* | *--* | *--* | *6.591* | *7.479* | *8.122* | *8.493* | *8.583* | *8.575* |
| 18 | **-1.307** | **2.268** | **4.444** | **6.305** | **7.244** | **7.872** | **8.289** | **8.510** | **8.663** |
| | *--* | *--* | *--* | *--* | *6.932* | *7.700* | *8.197* | *8.519* | *8.614* |
| 20 | **-1.730** | **1.709** | **3.874** | **5.820** | **6.847** | **7.559** | **8.051** | **8.341** | **8.550** |
| | *--* | *--* | *--* | *--* | *--* | *7.224* | *7.815* | *8.240* | *8.551* |

**Table 1. Theoretical (bold) and experimental (italic) binding energies per nucleon**

In Fig.1 the theoretical binding energies per nucleon along isotopic lines are plotted. Fig.2 shows the theoretical and experimental binding energies per nucleon for the N=Z nuclei. Also included in Fig.2 is the theoretical line obtained when one omits the variations of the coefficients that reflect the special features of light nuclei, and the potential well depth is kept constant at the maximum 67.5 MeV value (dashed line) in the range A=8 to A=40. The importance of taking into account the special features is thus clearly exhibited.

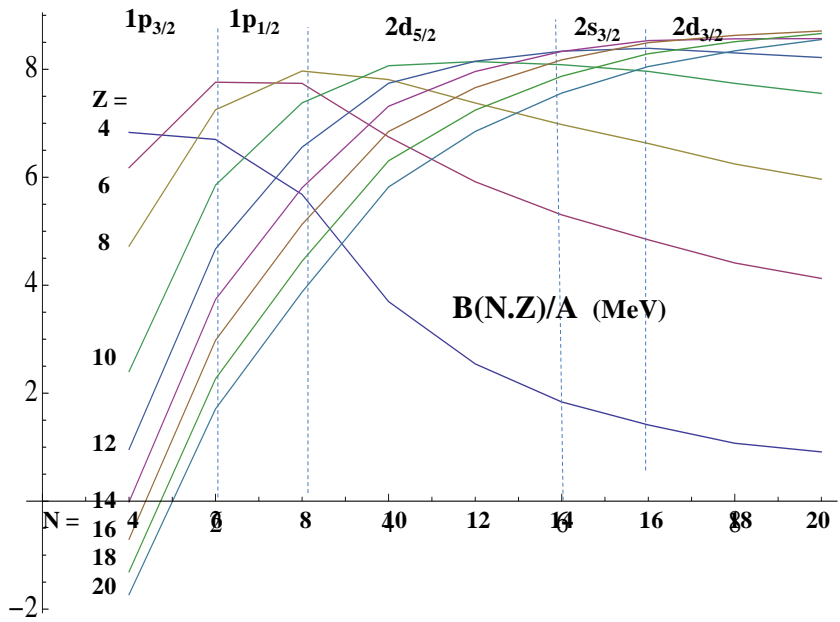

**Fig. 1 Theoretical binding energies per nucleon along isotopic lines**

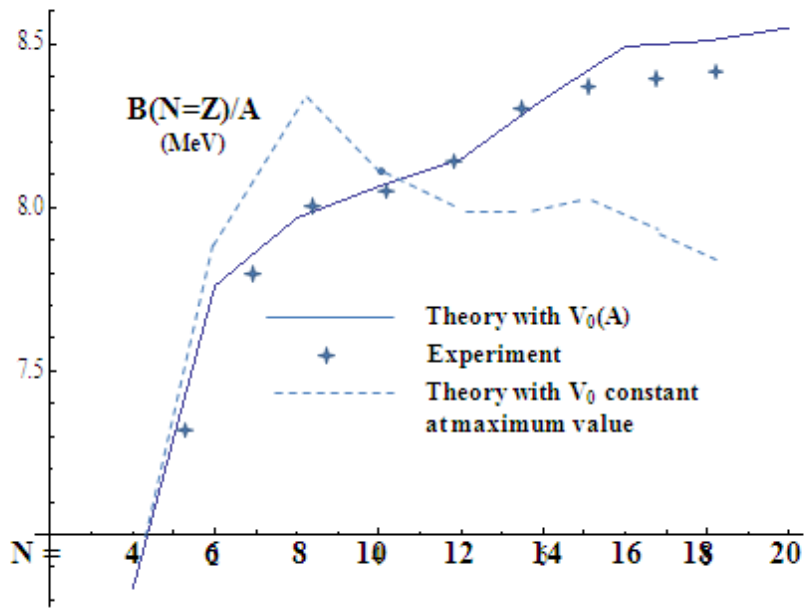

**Fig. 2. Theoretical and experimental binding energies per nucleon for N=Z nuclei**

## 4.2 Spin-orbit splitting and shell gaps

Fig. 3 exhibits the spin-orbit splitting of the neutron single particle levels and the shell gap for the Z = 4, 6, 8, 10 isotopes. It is seen that the splitting decreases as one goes towards neutron rich nuclei, an effect that has been already noted in light nuclei [27] as well as in heavier nuclei [28]. This is the consequence of the reduction of the harmonic oscillator frequency which follows from its asymmetry dependence as seen in Eqs. (11) and (12). In addition the gap between major shells (e.g., the difference in energy between the $0d_{5/2}$ and the $1p_{1/2}$ levels) also decreases.

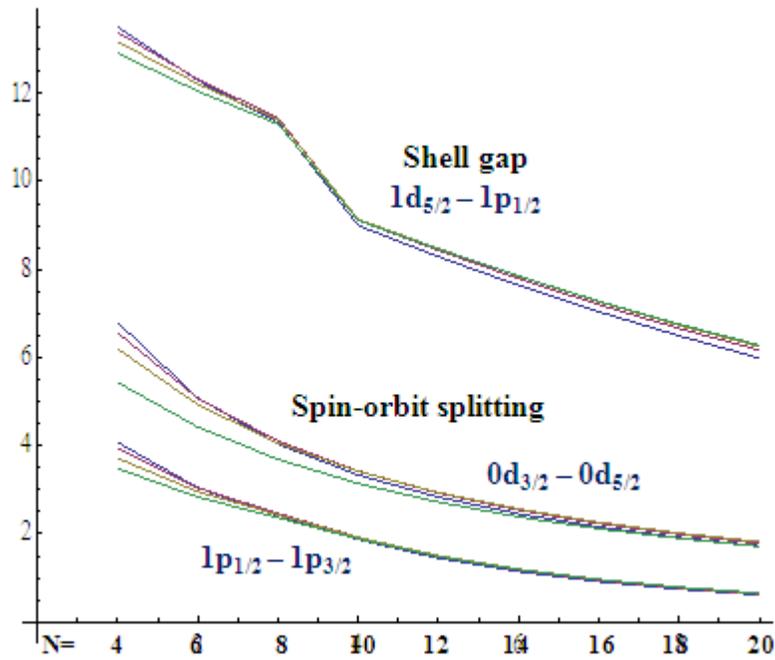

**Fig. 3** Spin-orbit splitting and shell gap along the isotopic lines Z=4, 6, 8 ,10 (MeV)

These two effects can give rise to configuration mixings when residual interactions are included that may result in shifting the magic numbers [6-9]. For example, the repulsion of the mixed configurations that may arise when the $1p_{3/2}$ and $1p_{1/2}$ levels get close to each other may result in N = 6 substituting N = 8 as a magic number.

## 4.3 Halo effect towards the neutron drip lines

The decrease in the neutron potential depth as (N-Z) increases (Eq.10) shifts all single particle levels towards the continuum, lowering the binding energies. Such effect can be reversed by a sudden increase of the nuclear potential range that increases the potential energy and lowers the kinetic energy, allowing those levels to be bound again into a halo configuration. Indeed, the halo isotopes set in after N=6 for Li and Be [16], and after N=14 for C with the two last neutrons occupying in this case the $2s_{1/2}$ level [9]. The data show clearly in both cases a sudden increase of the slope of the A dependence of the radius, or equivalently, a step like increase of the radial parameter $b_0$. Its origin is however different from the one arising along the valley of stability – the beginning of the occupation of a higher l orbital – as introduced above as feature c). As no change in the l orbital is involved, it can only arise from a readjustment of the self consistent nuclear potential to a larger size.

To test this, a calculation is carried out where the range parameter in the partially filled shell contribution (Eq. (A.8)) is increased beyond that of the filled shell (Eq. A.7). Fig.4 illustrates schematically these effects in the case of the 2s, 2d shell for the Z = 4, 6 and 8 isotopes, when the partially filled shell range is increased by 30 percent as N-Z exceeds two.

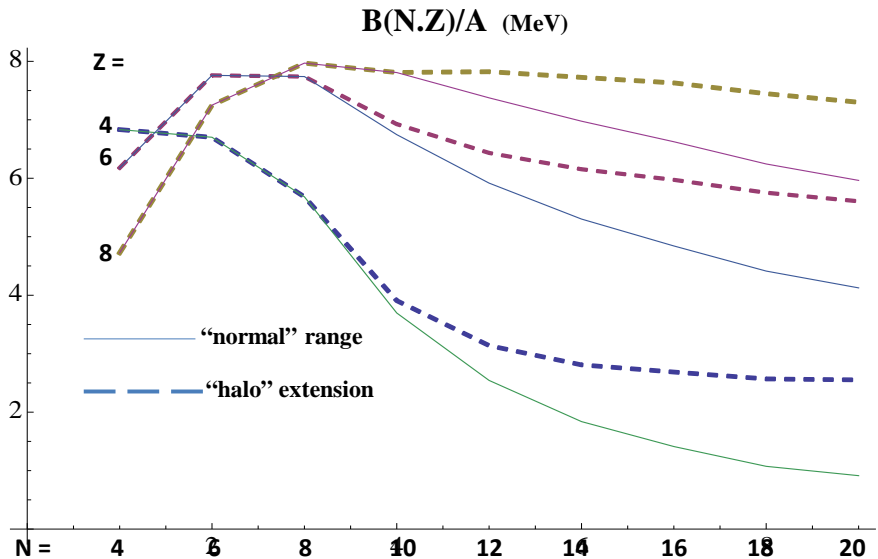

**Fig. 4. Binding energy shift by the halo effect towards the neutron drip line**

That the self consistent potential adjusts itself as needed to lower the ground state energy is not unknown, as witnessed by the onset of deformation away from the magic numbers along the valley of stability.

*4.4 Bethe-Weisszacker type formula*

As shown in Appendix C, the phenomenological shell model formula derived can be rewritten as a Bethe-Weisszacker type formula. However, there is no "surface" term correction, a feature linked to the use of a harmonic oscillator potential as better suited to the Gaussian shape of the mass distribution in light nuclei. With the parameter values used above, one obtains:

$$a_0 = 11.321 \text{ MeV} \; ; \; a_{0s} = 13.192 \text{ MeV}$$

$$a_1 = 0 \; ; \; a_{1s} = 0$$

$$a_2 = 8.637 \text{ MeV} \; ; \; a_{2s} = 3.388 \text{ MeV}$$

It is seen that the "volume" term is lower than the one from the usual B-W fit to all nuclei, compensating for the absence of a "surface" correction. Nevertheless this expression, together with the Coulomb term, yields a binding energy per nucleon of 8.836 MeV for $^{16}$O and 8.697 MeV for $^{40}$Ca, the doubly closed shell nuclei in the range considered. These are slightly above the calculate values (Table 1), which may be attributed to the approximations made. This may explain why larger differences arise in light nuclei when they are included in the usual B-W mass fits which include a surface correction [15].

## 5. Conclusion

In the present work it has been shown that a shell model mass formula needs to reflect the special features of light nuclei to provide an adequate analytic description of the ground state of light nuclei. The connection between the skin effect and the asymmetry term in optical model fits is established. The analytic structure derived also contains such effects as the decrease of the spin-orbit splitting of single particle levels and of the gap between major shells in neutron rich nuclei. Then it would be possible for configuration mixings induced by the residual interactions to open gaps at occupations different from the usual magic numbers. It also exhibits the possibility of increasing the binding energy by increasing the range of the partially filled shell, i.e., the onset of halo configurations. Furthermore, a quantitative agreement is obtained along the stability line with parameter values consistent with empirical evidence arising from the optical potential analysis of elastic scattering and from the spin-orbit splitting obtained from separation energies.

**Acknowledgments**

The authors would like to thank Dr. Mauricio Fortes for its assistance in the computational aspects of the present work.

**Appendix A. The effective mass approximation**

In the effective mass treatment [17, 19. 20] of a non local Schrödinger equation with a harmonic oscillator potential plus spin-orbit coupling, $W(r) = -W_0 + \frac{1}{2} m\omega^2 r^2 + C\,\mathbf{l.s}$, the energy level *nljm* has en energy given by:

$$e_{nljm} = -W_0 + \hbar\omega^*\left(n + \tfrac{3}{2}\right) - \mathcal{B}\left\{n(n+3) + l(l+1) + \tfrac{9}{2}\right\} - C\{j(j+1) - l(l+1) - \tfrac{3}{4}\}$$
(A.1)

where *n* is the principal quantum number, *l* the orbital quantum number ($l=0,2,...,n$ for *n* even and $l=1, 3,...,n$ for *n* odd) and $j=l\,(+/-)1/2$.

The H-F single particle energy contribution to the total energy is given by:

$$E_{nljm} = e_{nljm} - \tfrac{1}{2}<W(r)>_{nljm} =$$

$$-\tfrac{1}{2}W_0 + \mathcal{A}\left(n + \tfrac{3}{2}\right) - \mathcal{B}\left\{n(n+3) + l(l+1) + \tfrac{9}{2}\right\} - \left(\tfrac{C}{4}\right)\{j(j+1) - l(l+1) - \tfrac{3}{4}\}$$
(A.2)

with

$$\mathcal{A} = \left(\tfrac{1}{4}\right)\left[2 + \left(\tfrac{m*}{m}\right)\right]\hbar\omega^* \qquad \mathcal{B} = \left(\tfrac{1}{16}\right)\left[1 - \left(\tfrac{m*}{m}\right)\right](\hbar\omega^*)^2 \, / W_0$$

$$\hbar\omega^* = \sqrt{\tfrac{m}{m*}}\,\hbar\omega \tag{A.3}$$

The neutron or proton occupation number of a " *l* " sub shell is *2(2l+1)* taking into account the two spin projections, that of the *n* shell when summed over *l* values is *(n+1)(n+2)* and that of the major n shells summed up to $n_{max}= \eta$ is

$(1/3)(\eta+1)(\eta+2)(\eta+3)$. The partially filled shell $\eta+1$ contains $\alpha\,(\eta+2)(\eta+3)$, with $0 \leq a \leq 1$. Thus the total number of neutrons is given by:

$$N = (1/3)\,(\eta+1)(\eta+2)(\eta+3) + \alpha\,(\eta+2)(\eta+3) =$$

$$= (1/3)\,\{(\eta+2)^3 - (\eta+2)\} + \alpha\,\{(\eta+2)^2 + (\eta+3)\} \qquad (A.4)$$

This cubic equation can be solved to yield

$$\eta+2+\alpha \cong \{3N - \alpha(1-\alpha)(1-2\alpha)\}^{1/3}$$

$$+ (1/3)\{1-3\alpha(1-\alpha)\}\{3N - \alpha(1-\alpha)(1-2\alpha)\}^{-1/3} \qquad (A.5)$$

Similar expressions are valid for the proton number Z in terms of the corresponding proton maximum filled shell and occupation fraction of the partially filled shell. The contribution to the energy of a filled shell *n* is then

$$E(n) = \sum_l E_{nljm}[2(2l+1)] =$$

$$= \left[-\frac{1}{2}W_0 + \mathcal{A}\left(n + \frac{3}{2}\right) - (3/2)\mathcal{B}\{n(n+3)+3\}\right](n+1)(n+2) \qquad (A.6)$$

as the spin-orbit coupling with a constant coefficient gives a null contribution when both j sub shells are fully occupied. The sum over all filled shells up to a maximum *n* value $\eta$ is then:

$$E(\eta) = \sum_n E(n) =$$

$$\frac{1}{3}(\eta+1)(\eta+2)(\eta+3)\left[-\frac{1}{2}W_0 + \frac{3}{4}\mathcal{A}(\eta+2) - \frac{9}{10}\mathcal{B}\{(\eta+2)^2 + 1\}\right] \qquad (A.7)$$

The partially occupied major shell corresponds then to $n = \eta + 1$. Its contribution is taken as the average value $[1/(\eta+2)(\eta+3)]\,E(n=\eta+1)$ multiplied by the number of particles in the shell, written as $\alpha\,(\eta+2)(\eta+3)$ with $0 \leq a \leq 1$, to give:

$E(\eta+1)_{\text{part.filled.}} =$

$$= \alpha\,(\eta+2)(\eta+3)\left[-\frac{1}{2}W_0 + \mathcal{A}(\eta + 5/2) - \frac{3}{2}\mathcal{B}\{(\eta+1)(\eta+4)\}\right] \qquad (A.8)$$

As taking the average value overestimates the single particle energy at the beginning of the occupation of the shell, and underestimates it at the end, a correcting factor

$$f(\alpha,\eta) = -\{1.1 - 0.15\,\alpha\,(1 + 0.7\,\eta)\}$$

is introduced in eq. A.8

In addition there is a spin-orbit contribution given by:

$$- (C/4) \{\alpha (\eta+2)(\eta+3)(\eta+1)\} \qquad (A\ 9)$$

for $0 \leq a \leq (\eta+2)/(3\eta+3) + (2\eta)/(\eta+2)(\eta+3)$ as the $j = l + 1/2$ level is filled, followed by

$$- (C/4) (\eta+2)^2 (\eta+3)\{(2\eta+3)/3(\eta+1) + (2\eta)/(\eta+2)(\eta+3) - \alpha\} \qquad (A\ 9*)$$

to give a null contribution when both levels are filled ($\alpha = 1$). The intensity of the spin orbit interaction is $C = \lambda(\hbar\omega^*)^2 / 2mc^2$ where $\lambda$ is an adjustable parameter (Eq. 9').

## Appendix B. The crossing of a magic number

In addition to the skin effect related to the asymmetry of the optical potential, feature c) is taken into account as follows: in a harmonic oscillator potential, the rms radius of a shell n is given by

$$\langle r^2 \rangle_n^{1/2} = \left[ \left(\frac{\hbar}{m\omega}\right) \left(n + \frac{3}{2}\right) \right]^{1/2} \qquad (B.1)$$

The ratio between rms radius of the $n+1$ and $n$ shells is then $[1 + (n + 3/2)^{-1}]^{½}$. For the 1p and the (2d, 2s) shells this is of the order of 1.2. This can be parameterized by a progressive increase in the parameter $b_0$ as the shell is being filled, e.g., multiplying $b_0$ by a factor $(1 + a_4 \alpha \eta)$ where $\eta$ changes from zero to one on crossing the magic number 8 and $\alpha$ increases from 0 to 1 as the 2s-0d shell is filled. Correspondingly, a compensating factor $(1 + a_4 \alpha \eta)^{-3}$ is applied to $V_0$ (A).

## Appendix C. The Bethe-Weizsacker type mass formula

The ground state energy can be rewritten in a Bethe-Weizsacker type formula, using the approximations from Eq. (A.5):

$$(\eta+2+\alpha)_n \cong \{3N\}^{1/3} = (3A/2)^{1/3}\{[1+(N-Z)/A]\}^{1/3} \cong (3A/2)^{1/3}\{[1+(1/3)(N-Z)/A]\}$$

and

$$(\eta+2+\alpha)_p \cong \{3Z\}^{1/3} = (3A/2)^{1/3}\{[1-(N-Z)/A]\}^{1/3} \cong (3A/2)^{1/3}\{[1-(1/3)(N-Z)/A]\}$$

The summation over closed shells yields:

$$B(A,Z)/A = a_0 - a_1 A^{-1/3} - a_2 A^{-2/3} - a_3 A^{-1}$$
$$- a_{0s}\{(A-2Z)/A\}^2 - a_C [Z(Z-1)/b_0 A^{4/3}] \quad , \quad (C.1)$$

with

$$a_0 = 1/2\, V_0 \left(\frac{m}{m*}\right)_{av} - 1/8\,(3/2)^{\frac{1}{3}} \frac{(\hbar c)^2}{2mc^2} (1/b_0)^2\, (3/2)^{4/3} [2\left(\frac{m}{m*}\right)_{av} + 1] +$$

$$+ (27/320)(3/2)^{\frac{1}{3}} \left(\frac{(\hbar c)^2}{2mc^2}\right)^2 (1/b_0)^4 \left[\left(\frac{m}{m*}\right)_{av} - 1\right]/V_0$$

$$a_{0s} = 1/2\, V_1 \left(\frac{m}{m*}\right)_{av} \left[\left(\frac{m}{m*}\right)_{av} - \left(\left(\frac{m}{m*}\right)_{av} - 1\right)\left(\frac{V_1}{V_0}\right)\right]$$

$$- 1/3\,(3/2)^{\frac{1}{3}} \frac{(\hbar c)^2}{2mc^2} (1/b_0)^2 \left(\frac{m}{m*}\right)_{av} \left(\left(\frac{m}{m*}\right)_{av} - 1\right)\left(\frac{V_1}{V_0}\right)$$

$$a_1 = 0 \quad ; \quad a_{1s} = 0 \quad\quad\quad\quad (C.2)$$

$$a_2 = 1/3\,(3/2)^{1/3}\, V_0 \left(\frac{m}{m*}\right)_{av} - 1/8\,(3/2)^{\frac{1}{3}} \frac{(\hbar c)^2}{2mc^2} (1/b_0)^2\, (3/2)^{2/3} [2\left(\frac{m}{m*}\right)_{av} + 1] +$$

$$+ (9/160)\,(3/2)^{\frac{2}{3}} \left(\frac{(\hbar c)^2}{2mc^2}\right)^2 (1/b_0)^4 \left[\left(\frac{m}{m*}\right)_{av} - 1\right]/V_0$$

$$a_{2s} = 1/9(3/2)^{1/3}\, V_1 \left(\frac{m}{m*}\right)_{av} \left(\left(\frac{m}{m*}\right)_{av} - 3\left(\frac{m}{m*}\right)_{av} - 1\right)\left(\frac{V_1}{V_0}\right)$$

$$+ 1/12\,(3/2)^{2/3} \frac{(\hbar c)^2}{2mc^2} (1/b_0)^2 \left(\frac{m}{m*}\right)_{av} \left(\left(\frac{m}{m*}\right)_{av} - 1\right)\left(\frac{V_1}{V_0}\right)$$

There is no "surface" term. This arises from the fact that the total number of neutrons (protons) in closed shells is given by:

$$(1/3)(\eta+1)(\eta+2)(\eta+3) = (1/3)\{(\eta+2)^3 - (\eta+2)\}$$
$$\cong (3N) - (3N)^{1/3} \cong (3A/2) - (3A/2)^{1/3}$$

The contribution of the partially filled shells is here given by terms such as

$$\tfrac{1}{2}\,(3A/2)^{1/3}\,\{W_0{}^n\,[1+(1/3)(N-Z)/A]\,[\alpha(1-\alpha)\{1+4/3\alpha(1-2\alpha)]_n$$

$$+ W_0^p [1 - (1/3)(N-Z)/A] [\alpha(1-\alpha)\{1+ 4/3\alpha(1-2\alpha)]_p\}$$

that vanish at the beginning and the end of the shell, similarly to the spin orbit contribution.

**Corresponding author:** Mariano Bauer, Full Professor, theoretical nuclear physics, quantum mechanics, energy modeling; bauer@fisica.unam.mx